\documentclass[useAMS,usenatbib,usegraphicx]{mn2e}
\usepackage{pdflscape}
\usepackage{multirow}
\usepackage{natbib}
\usepackage{aas_macros}
\usepackage{amsmath}

\newcommand{\gasp}{\textsc{gasp2d}}
\newcommand{\Sersic}{S\'ersic }
\newcommand{\galaxyz}{\textsc{galaXYZ}}

\title[Deconstructing double-barred galaxies: bars]{Deconstructing double-barred galaxies in 2D and 3D. II. Two distinct groups of inner bars}
\author[A. de Lorenzo-C\'aceres et al.]{A. de Lorenzo-C\'aceres$^{1,2,3}$\thanks{E-mail:
adrianadelorenzocaceres@gmail.com}, J. M\'endez-Abreu$^{1,2,3}$, B. Thorne$^{1,4}$, and L. Costantin$^{5,6}$\\
$^{1}$School of Physics and Astronomy, University of St Andrews, North Haugh, KY16 9SS, Scotland, UK (SUPA)\\
$^{2}$Instituto de Astrof\'isica de Canarias (IAC), E-38205 La Laguna, Tenerife, Spain\\
$^{3}$Universidad de La Laguna (ULL), Departamento de Astrof\'isica, E-38206 La Laguna, Tenerife, Spain\\
$^{4}$Department of Physics, University of California, One Shields Avenue, Davis, CA 95616, USA\\
$^{5}$Centro de Astrobiolog\'ia (CSIC-INTA), Ctra de Ajalvir km 4, Torrej\'on de Ardoz, E-28850 Madrid, Spain\\
$^{6}$INAF Osservatorio Astronomico di Brera, via Brera, 28, 20159 Milano, Italy
}

\begin{document}
\date{Accepted ***. Received ***; in original form ***}

\pagerange{\pageref{firstpage}--\pageref{lastpage}} \pubyear{2018}

\maketitle

\label{firstpage}

\begin{abstract}
The intrinsic photometric properties of inner and outer stellar bars within 17 double-barred galaxies
are thoroughly studied through a photometric analysis consisting of:
i) two-dimensional
multi-component photometric decompositions,
and ii) three-dimensional statistical deprojections
for measuring the thickening of bars, thus retrieving their 3D shape. 
The results are compared with previous measurements obtained with the widely used analysis of
integrated light.
Large-scale bars in single- and double-barred systems show
similar sizes, and inner bars may be longer than outer bars in different galaxies.
We find two distinct groups of inner bars
attending to their in-plane length and ellipticity, resulting in a bimodal behaviour for the inner/outer bar length
ratio. Such bimodality is related neither to the properties of the host galaxy nor the dominant bulge,
and it does not show a counterpart in the dimension off the disc plane. The group of
long inner bars lays at the lower end of the outer bar length vs. ellipticity correlation, whereas
the short inner bars are out of that relation.
We suggest that this behaviour could be due to either a different nature of the inner discs
from which the inner bars are dynamically formed, or a different assembly stage for
the inner bars. This last possibility would imply
that the dynamical assembly of inner bars is a slow process taking several Gyr to happen.
We have also explored whether all large-scale bars are prone to develop an inner bar at some 
stage of their lives, possibility we cannot fully confirm or discard.
\end{abstract}

\section{Introduction}\label{sec:intro}
Double-barred galaxies are structurally complex systems due to 
the coexistence of several axisymmetric and non-axisymmetric components
within a disc galaxy, namely disc, outer bar, inner bar, and most likely a bulge
\citep[e.g.][hereafter Paper I]{Erwin2004, deLorenzoCaceresetal2019a}. Other
structures, such as spiral arms, inner discs, and lenses, may be present as well
\citep[see e.g.][]{deLorenzoCaceresetal2019b}.
Characterising stellar bars through their three main properties (length,
strength, and pattern speed) is therefore particularly difficult in the case
of double-barred galaxies.

While the measurement of the bar pattern speed usually requires spectroscopic data
\citep[see][ for an analysis of the bar pattern speed in a double-barred galaxy]{Corsinietal2003}, 
bar length and strength may be photometrically estimated on 
images with enough spatial resolution. Bars are considered triaxial ellipsoids
whose longest axis in the galaxy plane corresponds to the bar length.
The strength is a measurement of the prominence of a bar: 
it is  strong if it is long, massive, flat, elongated, 
and induces intense tangential forces, whereas weak bars 
are those which are small and with little gravitational influence 
on the rest of the galaxy. The most accurate way for estimating the 
bar strength is by means of the $Q_b$ parameter, introduced by 
\citet{ButaandBlock2001}. $Q_b$ is measured as the maximum value of the ratio 
between the tangential force and the mean axisymmetric radial force 
in a barred potential, and correlates well with other strength estimators, 
such as the bar ellipticity \citep[][]{Laurikainenetal2002}
or the parameterisation by \citet{AbrahamandMerrifield2000}.
The bar ellipticity is relatively simple to measure and it therefore is 
the most commonly used proxy for bar strength.

Photometric properties of bars have usually been  
derived through unsharp masking, ellipse fitting, or Fourier analysis 
(see \citealt{Aguerrietal98}, \citealt{Aguerrietal2000}, and \citealt{Erwin2004}, among others). 
The parameters obtained through these techniques are measured on the integrated light, where the overlapping 
with other bright structures such as the central bulge may be
affecting the results. Such contamination is particularly important in the case of a small inner bar
embedded in a double-barred system. Performing better suited two-dimensional (2D) multi-component
photometric decompositions of double-barred galaxies is complicated, and it has only 
been applied to a handful of individuals: two double-barred galaxies in 
\citet{MendezAbreuetal2017} and another two  in 
\citet{deLorenzoCaceresetal2019b}.

In Paper I we 
presented the most complete photometric analysis of double-barred galaxies ever 
performed on a sample of 17 individuals. It consists of a combination of 2D
photometric decompositions including a bulge, inner bar, outer bar, and
(truncated) disc, and a three-dimensional (3D) statistical deprojection
of bulges and bars thus retrieving their intrinsic 3D shape.
The main objective of this project is to use the accurate
photometric properties of double-barred galaxies to answer
four important questions still debated within the community, namely:
i) whether there exists a major incidence of disc-like bulges within
double-barred galaxies where secular evolution is assumed to take place in a very
efficient way; ii) whether inner bars form secularly after disc-like bulges
already present in barred galaxies; iii) whether inner bars are transient 
or long-lived structures;
and iv) whether all barred galaxies will develop an inner bar at some stage of their lives.

Questions i) and ii) are elaborated in Paper I, where we find that all galaxies
host a classical dominant bulge as indicated by the \citet{Kormendy1977} relation and their
intrinsic 3D shape 
\citep{Costantinetal2018b}. Such result poses the possibility that hosting 
a central hot component is a requirement for a barred galaxy to develop an 
additional inner bar.
In this second and last paper of the series, we present the photometric properties
of the inner and outer bars and compare them with previous results measured on
integrated galaxy light. We remark this is the first time the intrinsic photometric
properties of double bars are studied. We also address open questions iii) and iv), whose
relevance is put in context in the following.

\subsection{Formation of inner bars}
Two main formation mechanisms have been proposed for the case of double-barred
galaxies. The first scenario is a direct formation of the inner bar after gas inflow through the 
outer bar, as shown by various  simulations such as those from \citet{FriedliandMartinet93}, 
\citet{Helleretal2001}, \citet{EnglmaierandShlosman2004}, and \citet{Wozniak2015}. The gas is trapped in the $x_2$ orbits of the
outer bar, shaping a transient, gaseous inner bar; star formation is then 
triggered and a stellar inner bar eventually appears. The second possibility is
the formation of a double-barred system without the need of gas. 
\citet{DebattistaandShen2007} and \citet{Duetal2015} demonstrate that
inner bars may form as soon as a rapid-rotating component is present in the 
galaxy centre. Likewise single bars, which form dynamically out of a cold disc,
small-scale bars are formed out of small-scale discs. Disc-like
bulges, which are supposed to be frequent in barred galaxies
(but see the results shown in Paper I where we find a majority of classical bulges
in double-barred galaxies), may act
as the small-scale disc supporting the formation of an inner bar. 

Observational studies of inner bar formation are scarce in the literature, 
 with the notable exception of \citet{deLorenzoCaceresetal2012,
deLorenzoCaceresetal2013,deLorenzoCaceresetal2019b}.
These projects pursued the analysis of the stellar populations
and kinematics of double-barred galaxies. 
The results of these three articles, once combined, agree better with
 a formation through stellar redistribution supported 
by an underlying disc structure. All these works conclude notwithstanding that 
inner bars, once they are already formed, play
a very mild role in promoting secular evolution.

\subsection{Long-lived nature of inner bars}
Bars in general have been proposed to be transient structures that dissolve and reform
over time \citep[e.g.][]{BournaudandCombes2002, Wozniak2015}.  
Inflow processes can contribute to the bar destruction: it has been 
theoretically proved that large central mass concentrations may, 
together with the angular momentum exchange induced by the bar, 
dissolve it in rather short timescales \citep[$\sim$2\,Gyr; see for example][]{BournaudandCombes2002}.
On the other hand, some numerical simulations show that bars can be long-lived structures 
\citep[e.g.][]{DebattistaandSellwood2000, AthanassoulaandMisiriotis2002, MartinezValpuestaetal2006},
despite the combined effect of the gas 
flow and central mass concentrations \citep{Berentzenetal2007}.
Few observational results support this idea for the 
case of single bars \citep{SanchezBlazquezetal2011, Perezetal2017}.
Whether bars, or even inner bars, are long-lived structures is 
an important matter of debate and consensus has not been reached yet
(see e.g. \citealt{FriedliandMartinet93} and \citealt{Wozniak2015} for opposite results 
on the life time for inner bars).\\

The paper is organised as follows: the double-barred and comparison samples
are described in Sect.\,\ref{sec:data}. In Sect.\,\ref{sec:analysis} we summarise
the 2D and 3D analyses performed with \gasp\ and \galaxyz, respectively. 
The individual photometric properties of inner bars are presented in Sect.\,\ref{sec:ibprops}.
The results are discussed within the context of the formation and evolution of double-barred
galaxies in Sect.\,\ref{sec:discussion}.
Conclusions are wrapped up in Section \ref{sec:conclusions}.
A flat cosmology with $\Omega_{m}=$0.3, $\Omega_{\Lambda}=$0.7, and $H_0=$0.75 is 
assumed.
These are the same parameters adopted by
\citet{MendezAbreuetal2017}, whose work is used
for comparison throughout this paper.

\section[]{The samples of barred galaxies}\label{sec:data}
The sample of 17 double-barred galaxies photometrically analysed here
corresponds to all the barred galaxies with inner bars presented in 
\citet{Erwin2004} with available Sloan Digital Sky
Survey  \citep[SDSS;  ][]{Yorketal2000} images. While this first constraint
provided a list of 23 out of 50 objects, 6 of them 
(Mrk\,573, UGC\,524, NGC\,1068, NGC\,4303, NGC\,4321, and NGC\,4736)
were finally removed from the 
sample as either the SDSS spatial resolution is not enough for resolving their inner 
bars or the inner bar had been misclassified due to the 
presence of dust or other central components resembling elongated structures.

We use the $g'$-, $r'$-, and $i'$-band images of the SDSS Data Release
 9  \citep{Ahnetal12} for our photometric analysis. Besides the standard SDSS reduction, we
re-calibrate the images from nanomaggies to counts and refine the sky subtraction
\citep{Pagottoetal2017, Costantinetal2018b}.
Such additional treatment is  a requirement for our analysis, as explained in
Paper I.\\

\citet{MendezAbreuetal2017} analysed the photometric properties of 
a sample of galaxies from the CALIFA survey \citep{Sanchezetal2012}. The 2D photometric decompositions
performed by \citet{MendezAbreuetal2017} are analogous to those presented here:
they also applied the \gasp\ code to SDSS $g'-$, $r'-$, and $i'-$images.
404 galaxies are analysed in that work, among which there are 160 single-barred and
two double-barred hosts: NGC\,0023 and NGC\,7716, this last galaxy being in common
with the current sample. For the sake of completeness and to show the good agreement
between our results for the one galaxy in common, we have included the corresponding
measurements obtained by \citet{MendezAbreuetal2017} in the figures analysed throughout this paper.

\section[]{2D and 3D photometric analyses}\label{sec:analysis}

A detailed description of the procedures used for performing the 2D multi-component
photometric decomposition of the sample galaxies with \gasp\ \citep{MendezAbreuetal2008,
  MendezAbreuetal2014,  MendezAbreuetal2017}, as well as the 
3D statistical deprojection of the bulges, inner bars, and outer bars with
\galaxyz\ \citep{MendezAbreuetal2010,  Costantinetal2018a}, 
is presented in Paper I. For the sake of completeness
we summarise here the most relevant aspects of these methodologies.
Throughout this paper, we refer to the properties of the structures within the 
plane of the galaxy disc as \emph{in-plane} quantities, whereas
\emph{off-plane} properties correspond to the vertical direction, i.e. perpendicular
to the galaxy disc.

\subsection{2D analysis with {\sc gasp2d}}\label{sec:gasp}
\gasp\ uses a Levenberg-Marquardt algorithm for fitting
the 2D surface brightness distribution of each galaxy
with a combination of structural components, each of which is modelled with
a parametric mathematical function. In particular, we use 
\citet{Sersic68} profiles for the bulges, \citet{Ferrers77} profiles
for the bars, and single (double) exponential profiles for the
Type I (Type II or Type III) discs. 

We fitted the three ($g'$, $r'$, and $i'$) images for the 17 
double-barred galaxies, taking the $r'$-band image as benchmark
for the rest of the fits. Moreover, single-barred fits were also
performed in order to explore the effect  of the most common approach
of dismissing possible inner bars in photometric decompositions of barred
galaxies. The parameters describing all structural components present in every galaxy,
as well as their errors (computed in a Monte Carlo fashion by means of
mock galaxies),
are listed in Appendix\,A of Paper I. We remark again this is the first time
a 2D multi-component photometric decomposition analysis of 
a large sample of double-barred galaxies
including the inner bars is performed.\\

The Ferrers model used for the bars is characterised by two
shape parameters, namely $n_{bar}$ and $c$. 
$n_{bar}$ is related to the decay of the surface-brightness profile along the bar and
its value is highly correlated with  the bar length. $c$ indicates the
bar boxiness: $c=2$  represents a  perfectly elliptical bar while
$c>2$ and $c<2$ describe boxy and discy bars, respectively.
The standard procedure in photometric decompositions 
is to fix these parameters to their default values 
$n_{bar}=2$ and $c=2$ 
\citep[e.g.][]{Laurikainenetal2005, MendezAbreuetal2017}. With the aim of performing
an accurate analysis, we explore the space
of $n_{bar}$ and $c$ parameters and search for the most suitable value
for our inner and outer bars. We refer the reader to Paper I 
for an extensive description of this analysis. 

Our results indicate that the majority of both inner and outer bars
are well described with the default $c=2$, while the remaining galaxies
tend to host boxier bars. Regarding $n_{bar}$, all bars show
steeper profiles than the default $n_{bar}=2$. $n_{bar}$ and $c$
values for all our inner and outer bars are provided in the corresponding
tables presented in Paper I (Appendix\,A).\\

In summary, \gasp\ provides the shape of the isolated structural components
for every galaxy as projected onto the sky plane. The bar and bulge parameters
are then deprojected in order to retrieve the 
properties of each structure in-plane, i.e., within the galaxy plane.
To this aim, we perform the strict mathematical deprojection of an ellipse, 
following the equations shown in \citet{Gadottietal2007}. It is worth noting
that the Ferrers bar is not a classical but a generalized ellipse and therefore
some additional uncertainties are introduced by this deprojection. \citet{Zouetal2014}
perform an extensive study of deprojection uncertainties with simulated barred galaxies for
which the bars are not classical ellipses, finding that the mathematical
deprojection introduces a 10\% uncertainty of the same order of other 2D deprojections.
Note however
that the study from \citet{Zouetal2014} is focused on ellipse-fitting estimates; 
a similar study for parameters retrieved from photometric decompositions
is still lacking.

We remark here that, while deprojected in-plane position angles, ellipticities,
and sizes in physical units are used throughout this paper (as requested for a proper 
discussion and comparison between different galaxies),
observed results are listed in Appendix\,A of Paper I. For the sake of completeness, 
current Table\,\ref{tab:deprovalues} lists physical deprojected measurements (semi-major axes, 
effective radii, ellipticity, and position angle for the bars), 
as well as the physical disc effective radii for all sample galaxies as used throughout this paper.

\subsection[]{3D analysis with {\sc galaXYZ}}\label{sec:galaxyz}

Based on the properties of the structural components
obtained through 2D photometric decompositions,
\galaxyz\ performs a statistical deprojection
in order to retrieve the shape off the galaxy plane, thus
providing the full 3D morphology of the isolated structures. For \galaxyz\
to be properly applied, three conditions must be fulfilled: 
i) the structure under study is well modelled as a triaxial ellipsoid; ii)
the galaxy disc is an oblate spheroid; and iii) disc and structure
share the same centre. This technique has been successfully used with
bulges \citep{Costantinetal2018a} and large-scale bars \citep{MendezAbreuetal2018b},
and its pioneering application to bulges, inner bars,
and outer bars within double-barred systems was presented in Paper I.

While it is known that bars may develop vertical instabilities which may
give rise to vertical-extended components such as box/peanut structures
\citep[e.g.][among others]{MartinezValpuestaetal2006}, 
we refer the reader to \citet{MendezAbreuetal2018b} for a demonstration
of how the parameters retrieved with \galaxyz\ correspond to the
thin part of the bars. Central vertical components such as box/peanuts
do therefore not affect the results.

The outcome of \galaxyz\ is the joined probability distribution
function (PDF) for the in-plane ($B/A$)
and off-plane ($C/A$) axis ratios of the structures under study. A summary of the
mathematical equations used in this analysis is presented 
in \citet{Costantinetal2018a}. In Paper I
we show that the deprojected in-plane axis ratios derived with \gasp\ match
very well those obtained using \galaxyz. Table\,2 of Paper I lists 
the results for the 17 sample galaxies of this project.

\section[]{Photometric properties of bars within double-barred systems}\label{sec:ibprops}
Here we explore for the very first time
the individual photometric properties of inner and outer bars in double-barred galaxies, which are 
worth comparing with previous values and conclusions from analyses with the most extensively used ellipse fits over
integrated-light images.

\subsection[]{Lengths of inner and outer bars}\label{sub:comparison}

\begin{figure}
 \vspace{2pt}
 \includegraphics[angle=0., width=0.45\textwidth]{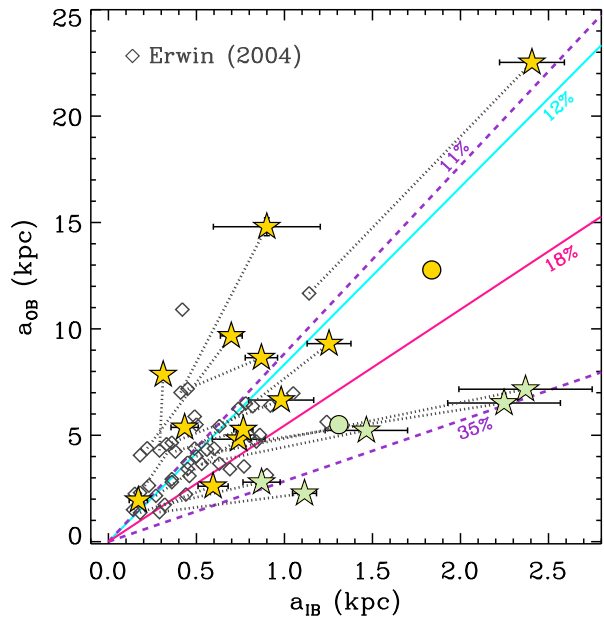}
 \caption{Comparison between the deprojected outer ($a_{\rm OB}$) 
and inner ($a_{\rm IB}$) bar semi-major axes obtained from \gasp\ 
in $r'$-band for our double-barred sample (stars) and the two double-barred galaxies 
included in the analysis of \citet[][circles]{MendezAbreuetal2017}. 
Yellow symbols represent double-barred galaxies
with $a_{\rm IB}/a_{\rm OB}<0.23$, 
whereas green symbols correspond to  $a_{\rm IB}/a_{\rm OB}>0.23$.
Grey diamonds show the deprojected $a_e$ 
ellipse-fitting measurements for 44 double-barred galaxies
from the catalog of \citet[][six galaxies were excluded due to misclassification
or lack of resolution, 
see Sect.\,\ref{sec:data} for details]{Erwin2004}. Grey dotted lines connect the values for our sample galaxies
with their corresponding ellipse-fitting measurements. 
We obtain an average $a_{\rm IB}/a_{\rm OB}$ ratio of 18\%, indicated with a solid magenta line.
For comparison, the previously computed 12\% ratio derived by
\citet{ErwinandSparke2002} is indicated with a solid
cyan line. The two purple dashed lines show either
ratios when a bimodal distribution is taken into account (11\% and 35\%).}
 \label{fig:lengthcomparison}
\end{figure}

\begin{figure*}
 \vspace{2pt}
 \includegraphics[angle=0., width=.95\textwidth]{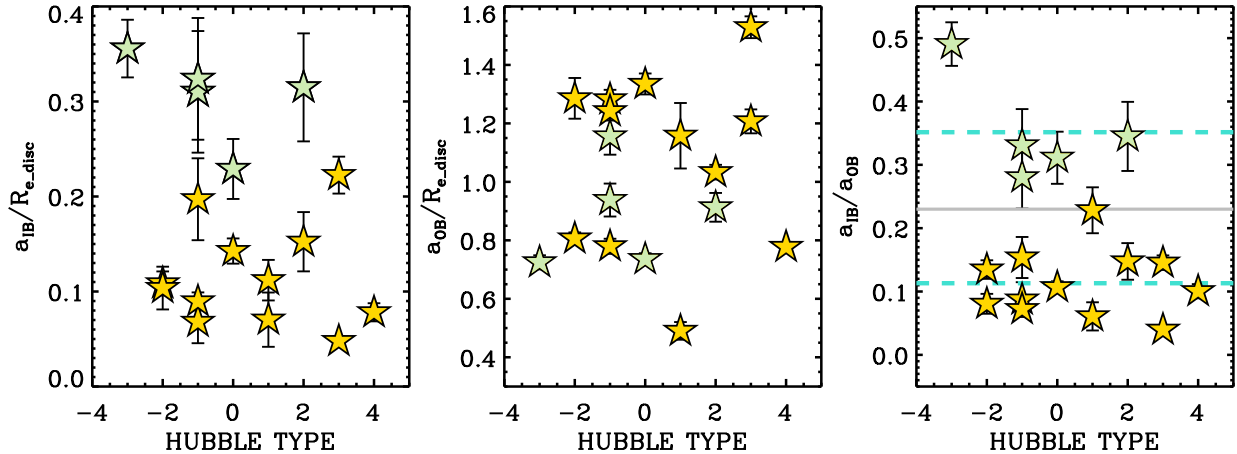}
 \caption{Deprojected inner (left panel) and outer (middle panel)
bar semi-major axes (in units of the disc effective radius of every galaxy), as well as their ratio (right panel), with respect
to the morphological type of the host galaxies. Bar lengths correspond to \gasp\ measurements
on the $r'$-band images. Yellow stars represent double-barred galaxies
with $a_{\rm IB}/a_{\rm OB}<0.23$, 
whereas green stars correspond to $a_{\rm IB}/a_{\rm OB}>0.23$.
In the right panel, the dashed cyan lines indicate the 
two bar length ratios obtained in this work (11\% and 35\%), whereas the solid grey line indicates the
demarcation ratio (23\%).}
\label{fig:lengthvsht}
\end{figure*}

\begin{figure*}
 \vspace{2pt}
 \includegraphics[angle=90., width=0.8\textwidth]{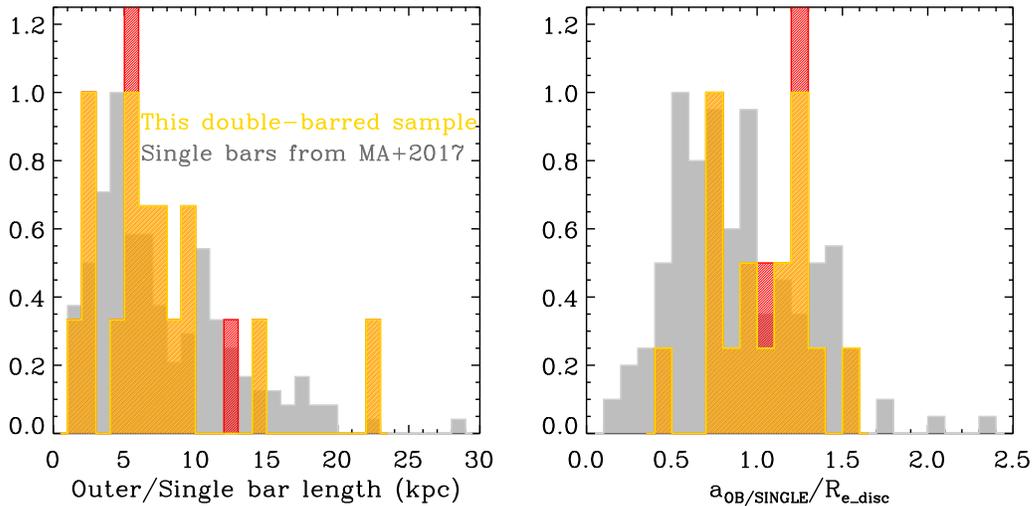}
 \caption{Distribution of deprojected large-scale bar semi-major axes: outer bars within double-barred galaxies probed 
in this work are plotted in yellow, while the CALIFA single bars analysed in \citet{MendezAbreuetal2017}
are shown in grey. Both distributions are normalised to their maximum value for the sake of comparison.
The red segments show how the distribution studied in this paper is modified when including the two 
double-barred galaxies from \citet{MendezAbreuetal2017}. The left panel shows physical sizes while the right panel shows bar lengths in units
of disc effective radius for every galaxy, thus preventing biases due to the size of the whole galaxy.}
\label{fig:barssingledouble}
\end{figure*}

Figure \ref{fig:lengthcomparison} shows the deprojected inner and outer bar lengths as 
measured with photometric decompositions. Inner bars are rather small systems
with physical lengths ranging from 0.17\,kpc to 2.4\,kpc (semi-major axis). 
Outer bars can be as short as 1.9\,kpc, meaning that
some galaxies host outer bars shorter, or of the same length, than inner bars in other galaxies.

Ellipse-fitting measurements from
\citet{Erwin2004} are shown in Fig.\,\ref{fig:lengthcomparison}
for comparison. 
Photometric decompositions provide systematically longer bar
lengths with respect to ellipse-fitting techniques, as already observed
 by \citet{Gadotti2011}. This trend is expected as our
photometric decompositions measure the whole extent of the Ferrers profile, i.e. up to
when its contribution to the total galaxy light drops to zero.
It therefore represents the actual individual bar length.
On the contrary, bar-length measurements from ellipse fits recover estimates of the
extent to which signatures of the bar presence
on the total integrated light (i.e. the image) are found: when the surface brightness 
of the bar is dropping to zero, its effects over the galaxy isophotes become negligible.

\citet{Erwin2004} provides four different ellipse-fitting estimates of the bar length. 
We note that the values shown in
Fig.\,\ref{fig:lengthcomparison} correspond to the lower limit $a_e$, measured as 
the radius of maximum ellipticity within the bar. We chose this parameter
because it is the only one available in \citet{Erwin2004} for all our sample galaxies. We remark that, 
although other measurements such as $a_{\rm 10}$ and $a_{\rm min}$ 
\citep[we refer the reader to][for details on how these estimates are derived]{Erwin2004} 
provide larger bar length values,
these are still shorter
than the individual sizes derived from photometric decompositions.

While bars measured through photometric decompositions are always longer than 
 ellipse-fitting estimates, \citet{Gadotti2011} finds an approximated match for both quantities when 
the effective radii of bars are considered instead of their full length. 
To make a similar comparison, we have calculated the effective radii of our Ferrers profiles for the inner bars
(see Table\,\ref{tab:deprovalues}).
We find a good agreement between the inner bar effective radii and $a_e$ measurements from \citet{Erwin2004},
with a median deviation of 18\% for the whole sample
that can be as low as 2\% for some galaxies. This result further supports the fact that photometric analyses of integrated light
focus on the brightest regions of bars, particularly inside their effective radii. 
Such conclusion is robust against different parametric functions of the bars, since \citet{Gadotti2011} uses
\Sersic instead of Ferrers profiles.

Figure \ref{fig:lengthcomparison} also shows a slight correlation so galaxies with shorter outer
bars tend to host shorter inner bars, although a non-negligible dispersion is present.
In average, our measured inner bars have 18\% the in-plane
size of outer bars, in contrast with the 12\% ratio previously reported by 
\citet{ErwinandSparke2002} with ellipse fits.
By inspecting the Spearman factor of this ratio, we
find the global correlation is indeed very weak ($\rho$=0.34 with a significance of 0.18). 
Fig.\,\ref{fig:lengthcomparison} suggests that a bimodal behaviour of the bar length ratio may hold.
If such bimodality is taken into account, two strong trends between the bar sizes are found.
We infer that, when galaxies are separated according to a bar length ratio 
above and below $a_{\rm IB}/a_{\rm OB}$=0.23,
a clear correlation of $a_{\rm IB}/a_{\rm OB}$=35\% ($\rho$=0.90 and a significance of 0.04)
and a milder, although still significant relationship $a_{\rm IB}/a_{\rm OB}$=11\% 
($\rho$=0.65 and a significance of 0.02), appear. The global mean ratio and the
two trends are slightly transformed into 17\%, 
32\%, and 11\%, respectively, when effective radii instead of full bar lengths are considered.

The origin of the possible bimodality in the size ratios is not known but it is not related to 
the Hubble type of the
host galaxy, as demonstrated in the right panel of 
Fig.\,\ref{fig:lengthvsht}. We have also explored whether it could be related to the central bulge,
but no global correlations between bar length and bulge parameters have been found
(see Fig.\,5 in Paper I).
The left and central panels
of Fig.\,\ref{fig:lengthvsht} show how the bimodality in
$a_{\rm IB}/a_{\rm OB}$ is mainly driven by the inner bars, as those with $a_{\rm IB}/a_{\rm OB}>$0.23 are 
the longest inner bars in the sample.\\

Numerical works exploring the formation and evolution of double-barred galaxies
predict the size evolution of a two-bars system. Numerical simulations by \citet{Wozniak2015}
form a double bar in a two-steps process: a  transient inner bar is first created 
with a very small inner/outer bar length ratio of just 5\%; after the dissolution
of this first inner bar,
a second, long-lived inner bar with $a_{\rm IB}/a_{\rm OB}$=0.275 is created, and it
rapidly evolves in size and eventually gets $a_{\rm IB}/a_{\rm OB}\sim$0.15.
Other authors provide similar measurements, such as \citet[][26\%]{FriedliandMartinet93}
and \citet[][from 10\% to 16\%]{Duetal2015}. They all lay within the range covered by the
ratios measured in this work and shown in the right panel of Fig.\,\ref{fig:lengthvsht}. 

Fig.\,\ref{fig:lengthvsht} also shows four galaxies with $a_{\rm IB}/a_{\rm OB}\geq$0.3.
\citet{SahaandMaciejewski2013} report on the spontaneous formation of double-barred systems with 
length ratios as large as 50\%, starting with a dark-matter-dominated model that includes a
disc, classical bulge, and halo, but no gas. We note that 
this simulation forms both bars rather simultaneously, 
against the available observational evidence 
\citep[e.g.][]{deLorenzoCaceresetal2013,deLorenzoCaceresetal2019b}. Still, simulations
must reproduce the formation of double bars with $a_{\rm IB}/a_{\rm OB}\geq$0.4, since 
observational evidence for such systems has been provided not only by this work but also
by other authors \citep[e.g. the deprojected $a_{10}$ bar lengths 
for NGC\,3358 in][account for $a_{\rm IB}/a_{\rm OB}=$0.43]{Erwin2004}.

We note here that the highest bar length ratio in Fig.\,\ref{fig:lengthvsht}
corresponds to NGC\,3941 ($a_{\rm IB}/a_{\rm OB}$=0.49). For this galaxy, \citet{Erwin2004}
measures a deprojected bar length ratio of $\sim$0.2 from ellipse fitting. 
When our bar full lengths are translated into effective radii, 
we obtain a similar inner bar size (as expected) but a shorter outer bar radius than from ellipse fitting,
thus resulting in a still high $R_{\rm e\_IB}/R_{\rm e\_OB}=$0.43 for this galaxy. We remark again
ellipse fitting and photometric decompositions are different techniques and only the latest is able to isolate
the galaxy structures thus providing their true parameters.\\

Figure\,\ref{fig:barssingledouble} shows the distribution of deprojected large-scale bar lengths for
the double-barred galaxies of this sample (i.e., outer bars) and the single-barred galaxies
from \citet{MendezAbreuetal2017}, as 
it has been argued that outer bars in double-barred
systems are longer than single bars \citep[see e.g.][]{Erwin2011}. 
We remind the reader that both samples of galaxies 
have been analysed in an analogous way with \gasp\ and therefore they represent
the best double/single-barred galaxies pair of samples to probe this.
Moreover, we have performed the deprojection of the bar lengths
for the galaxies of \citet{MendezAbreuetal2017} following the same recipe than for our double-barred sample. 
As seen in the left panel of Fig.\,\ref{fig:barssingledouble}, both distributions peak at approximately
5\,kpc, with median values of 6.3\,kpc and 6.5\,kpc for the single and double bars, respectively. 
The bulk of double-barred galaxies
host outer bars which populate the central parts of the distribution for single-barred galaxies,
although there are two double-barred systems with quite large bars as well.

The right panel of Fig.\,\ref{fig:barssingledouble} shows analogous measurements, but bar lengths
are normalised by the effective radius of the disc included in Table\,\ref{tab:deprovalues}. This is done 
in order to prevent biases due to trends inherent 
to differences in the galaxy hosts, not related to the nature of bars. Note that, indeed, 
no galaxy with particularly large bar stands out. This indicates that the two largest outer bars 
are hosted by also-large galaxies. The distribution for double bars overlaps with that for single
bars and they have median values of 1. and 0.8, respectively.
We must highlight the different statistics accounted for in the two samples 
(160 single-barred galaxies in CALIFA against the 17+2 probed in this work).
The conclusion is notwithstanding that outer bars within double-barred systems
are not systematically longer than single bars.

\subsection[]{Bar length versus morphological type}\label{sub:ht}

For large-scale bars (i.e., single bars or outer bars in double-barred systems), it has been found
that they tend to be shorter in later-type galaxies 
\citep[e.g.,][]{Martin95, Laurikainenetal2002, Erwin2005, Aguerrietal2009}.
With the aim at exploring this trend on inner bars as well, Fig.\,\ref{fig:lengthvsht} shows
the individual bar lengths measured from \gasp\ versus the morphological type provided by
\citet[][see Table\,1 in Paper I]{Erwin2004}.
No correlations are found for either the inner or outer bar.
This lack of trend stands for
physical bar lengths (not normalised by disc effective radius).

Our sample covers up to Sbc types, with very few individuals in our latest-type
regime. The decreasing 
(large scale) bar length behaviour mentioned before holds for Sc-Sd galaxies, whereas 
Sb-Sbc galaxies have been found to span a large range in bar length
\citep[see Fig.\,11 in ][]{Erwin2005}, in agreement with our measurements for outer bars.
The slight decrease of outer-bar lengths for Hubble types $<-1$ is also
in agreement with previous findings. We also remark that recent results from \citet{Erwin2019}
indicate that the relation between bar length and morphological type
is indeed weak and most likely driven by the correlation between morphological type and galaxy mass.

\subsection{Bar position angles}\label{sub:barpas}

\begin{figure*}
 \vspace{2pt}
 \includegraphics[angle=0., width=0.8\textwidth]{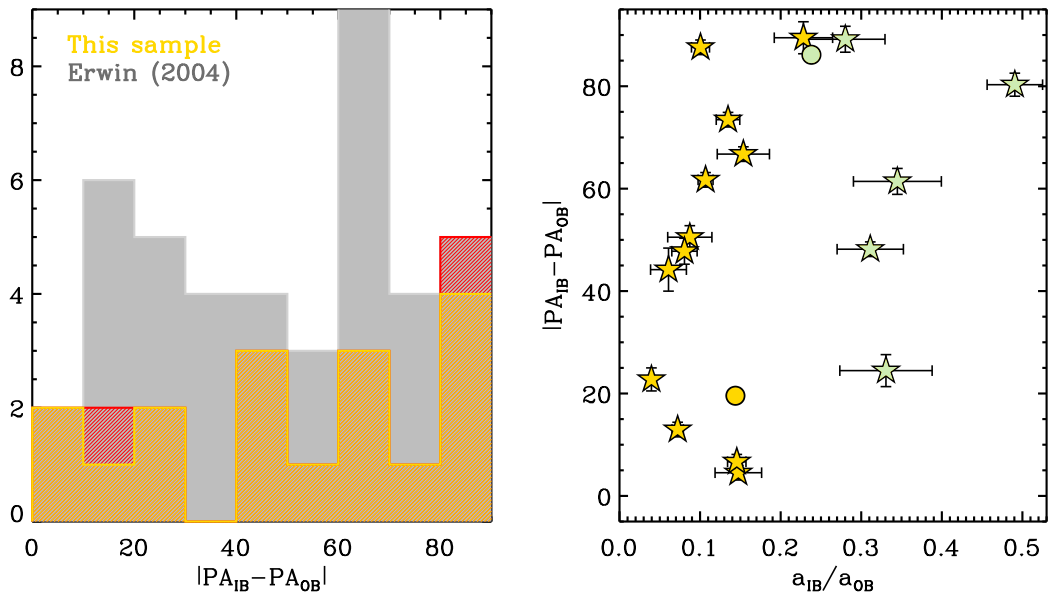}
 \caption{Left panel: distribution of the difference between the deprojected position angles of the inner and outer bars 
for the double-barred galaxies analysed with photometric decompositions (this sample; yellow) and with ellipse fitting 
\citep[][grey]{Erwin2004}. The red segments show how the distribution studied in this paper is modified when 
including the two double-barred galaxies from \citet{MendezAbreuetal2017}. The wide coverage confirms the random orientation 
between the two bars. Right panel: difference in the inner and outer bar position angles versus the bar length ratios 
for the double-barred galaxies of this sample (stars) and the two double-barred galaxies (circles) included in 
\citet{MendezAbreuetal2017}. Yellow symbols represent double-barred galaxies with $a_{\rm IB}/a_{\rm OB}<0.23$, 
whereas green symbols correspond to $a_{\rm IB}/a_{\rm OB}>0.23$. No particular trend between the two quantities is found.}
\label{fig:barpas}
\end{figure*}

Two relevant pieces of information can be inferred from the position angles
of the two bars within a double-barred system. First, the random relative
orientation of the bars has been used as demonstration of their independent
rotation, since a preferred relative angle would be expected otherwise
\citep[e.g.][]{FriedliandMartinet93}. 
Left panel in Fig.\,\ref{fig:barpas} shows the relative position angles measured 
with photometric decompositions in this work together with the ellipse-fitting results
from the larger sample of \citet{Erwin2004}. As expected, the distribution
covers the whole range and agrees with the two bars having independent pattern speeds.
This is a robust result since the measurements of the position angles have been corrected
for the galaxy inclination.
A proper confirmation of this result has been performed
for NGC\,2950 via a Tremaine-Weinberg analysis by \citet{Corsinietal2003}
and for a handful of other double-barred galaxies with indirect techniques \citep{Fontetal2014}.\\

Second, numerical simulations predict the formation and dynamical evolution of a double-barred
system and the bar position angles may therefore be used to analyse the goodness
of the predicted scenarios. It is widely accepted that both bars 
grow in length and strength during 
their lifetimes, and their evolution depends on their relative position, 
inner bars being longer (and axis ratios being higher) when both bars are perpendicular.
This trend is indeed found in many numerical works such as those of 
\citet[][and references therein]{Duetal2015, Wozniak2015, SahaandMaciejewski2013},
and it finds its physical explanation in the orbital analysis of double
bars performed by \citet{MaciejewskiandSparke2000} and \citet{MaciejewskiandSmall2010}.
The pattern speed also oscillates so inner bars rotate slower when both bars are perpendicular.
It is therefore sensible to state that it is most likely to find orthogonal  
than almost-parallel double bars, and a correlation between the bar-length ratios 
and the relative position angles is expected. 

This hypothesis is
not supported by the results shown in the right panel of Fig.\,\ref{fig:barpas}, in which
the distribution of double bars along the relative position angles is 
rather homogeneous and there is no noticeable correlation with respect to 
the bar axis ratios. We must note here that the oscillating amplitudes, strength, and pattern
speeds behaviour found in the simulations applies to settled-up bars, while 
the proper formation process before is quite chaotic. For example, in \citet{SahaandMaciejewski2013}
inner bars are born as very slow structures that progressively speed up and eventually rotate faster
than outer bars.

\subsection{Bar ellipticities and 3D shapes} \label{sub:barellipticities}

\begin{figure*}
 \vspace{2pt}
\centering
 \includegraphics[angle=0., width=0.8\textwidth]{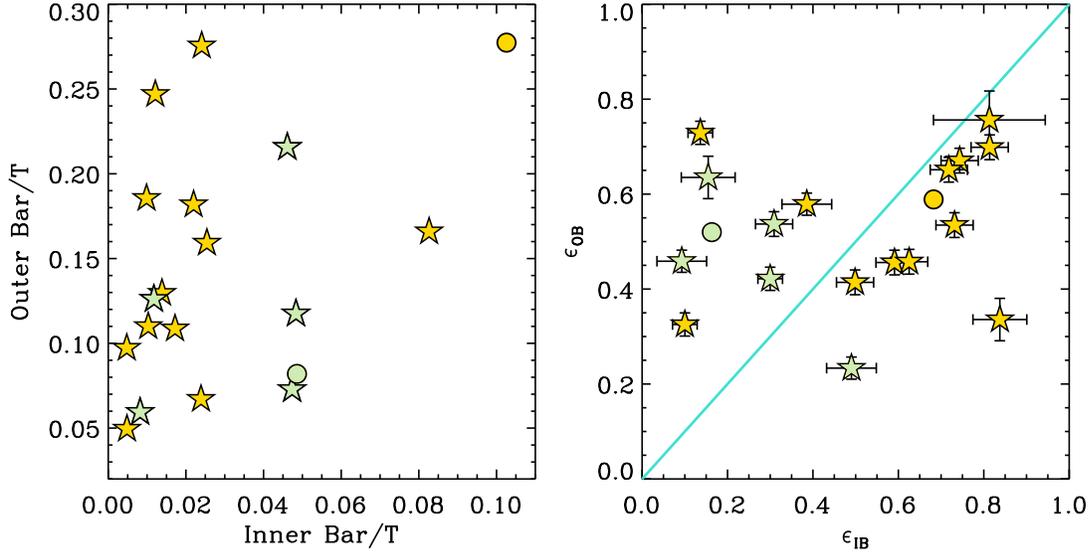}
 \caption{Inner and outer bar contributions to the total galaxy light (left panel) 
and deprojected ellipticities (right panel) obtained with \gasp\ 
in the $r'$-band for our double-barred sample (stars)
and the two double-barred galaxies included in the analysis of \citet[][circles]{MendezAbreuetal2017}.
Yellow symbols represent double-barred galaxies
with $a_{\rm IB}/a_{\rm OB}<0.23$, 
whereas green symbols correspond to $a_{\rm IB}/a_{\rm OB}>0.23$.
The cyan solid line in the right panel highlights
the 1:1 relationship.}
 \label{fig:comparisonbarprops}
\end{figure*}

\begin{figure}
 \vspace{2pt}
\centering
 \includegraphics[bb=100 20 620 540, angle=0., width=0.5\textwidth]{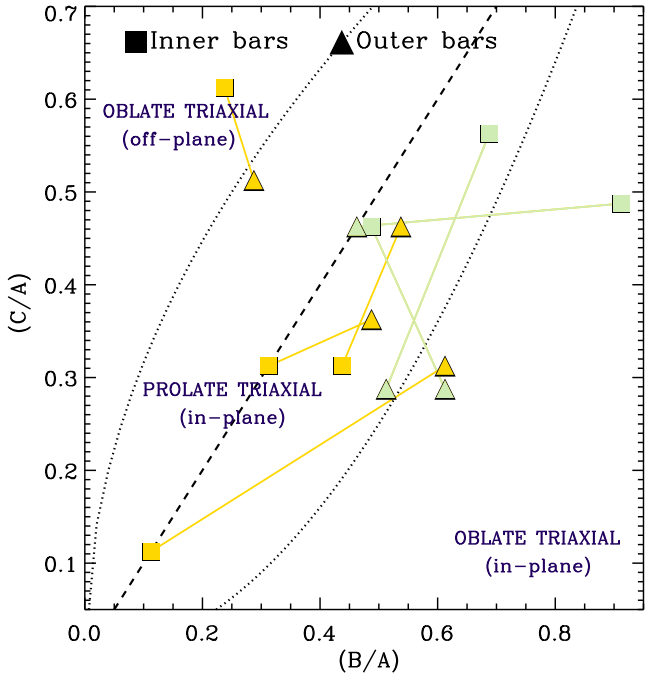}
 \caption{Intrinsic semi-major axis ratios for the inner (squares) and outer (triangles) bars obtained with \galaxyz\ for
the seven double-barred galaxies with 1$\sigma$ uncertainties $<$0.5 in any of the involved parameters. 
$A$ is the bar longest semi-major axis
in the galaxy plane; $B$ is the bar shortest semi-major axis in the galaxy plane; and
$C$ is the bar semi-major axis perpendicular to the galaxy plane. Yellow symbols represent double-barred galaxies
with $a_{\rm IB}/a_{\rm OB}<0.23$, 
whereas green symbols correspond to galaxies with $a_{\rm IB}/a_{\rm OB}>0.23$.
Structures for the same galaxy are connected with a solid line.
The regions where oblate triaxial and prolate triaxial structures lay are indicated.
To guarantee a clear presentation of the results, error bars are not shown in this plot;
they can be found in Table 2 of Paper I. }
 \label{fig:comparisonbar3D}
\end{figure}

The right panel of Fig.\,\ref{fig:comparisonbarprops} shows the deprojected ellipticity values for 
our sample of inner and outer bars. Both bars span almost the full range of ellipticities. It is
particularly noticeable that inner bars can be as round as $\epsilon\sim$0.1 and as elongated as
$\epsilon\sim$0.85. Outer bars are constrained to the slightly narrower regime $\epsilon\in$ [0.23,0.76].

We remark that the ellipticities shown in Fig.\,\ref{fig:comparisonbarprops} 
have been corrected for galaxy inclination. \citet{Erwin2004} does not provide deprojected measurements
of the ellipticities. For this reason, we do not overplot his results in Fig.\,\ref{fig:comparisonbarprops},
but we note that the ellipse fitting provides systematically rounder bars in projection, as expected: 
contamination from the bulge
light generates rounder isophotes in the central regions.
\\

Ellipticity is often used as a proxy for bar strength \citep{Laurikainenetal2002}. 
The results shown in Fig.\,\ref{fig:comparisonbarprops} indicate that 
double-barred galaxies can host either stronger outer bars than inner bars, or the opposite.
It is worth noting that longer inner bars (coloured in green) tend to be rounder 
(i.e. weaker) than outer bars, even though
they correspond to rather prominent structures with respect to the total galaxy light, as
observed in the left panel of Fig.\,\ref{fig:comparisonbarprops}.
We must however note that the vertical shape of a barred structure also 
affects the gravitational potential it introduces. 
The 3D statistical deprojection presented here allows us to take into account not 
only the deprojected ellipticity (or $B/A$ axis ratio), but also the off-plane
axis ratio $C/A$ when studying the  influence of the bar on the galaxy. Thin bars 
have a stronger effect than thick bars and they may, for example, trigger
star formation or promote secular evolution in a more efficient way.

Figure\,\ref{fig:comparisonbar3D} shows the in-plane and off-plane axis ratios
for the outer and inner bars of the double-barred sample. Only those galaxies
for which the uncertainties in both axis ratios for the two bars are less than 0.5 are
shown (we refer the reader to Paper I for more details on this threshold).
This analysis is therefore restricted to seven out of 17 galaxies, with five of the inner bars mostly being
prolate triaxial ellipsoids whereas only the other two are consistent with an oblate shape. Attending 
to the two axis ratios available, 
two galaxies host clearly stronger outer bars than inner bars
(i.e., outer bars are thinner and more elongated than inner bars) and three galaxies show stronger inner bars.
The remaining two galaxies show opposite behaviours for the two proxies under use.
No clear trend is therefore found for the bar strength
when not only the ellipticity but also the vertical thickness is analysed.

\section[]{Discussion: formation and evolution of double-barred galaxies}\label{sec:discussion}
We discuss here the previous results within the context of the two major 
questions brought up in Sect.\,\ref{sec:intro} concerning the formation and evolution
of double-barred galaxies: iii) whether inner bars are transient 
or long-lived structures;
and iv) whether all barred galaxies will develop an inner bar at some stage of their lives.
Other related issues, such as at which precise stage of their evolution
we are witnessing inner bars, are addressed too.

\subsection{Assembly of outer and inner bars}\label{sub:discussionsettle}

\begin{figure*}
 \vspace{2pt}
 \includegraphics[angle=0., bb=54 15 680 254, width=0.9\textwidth]{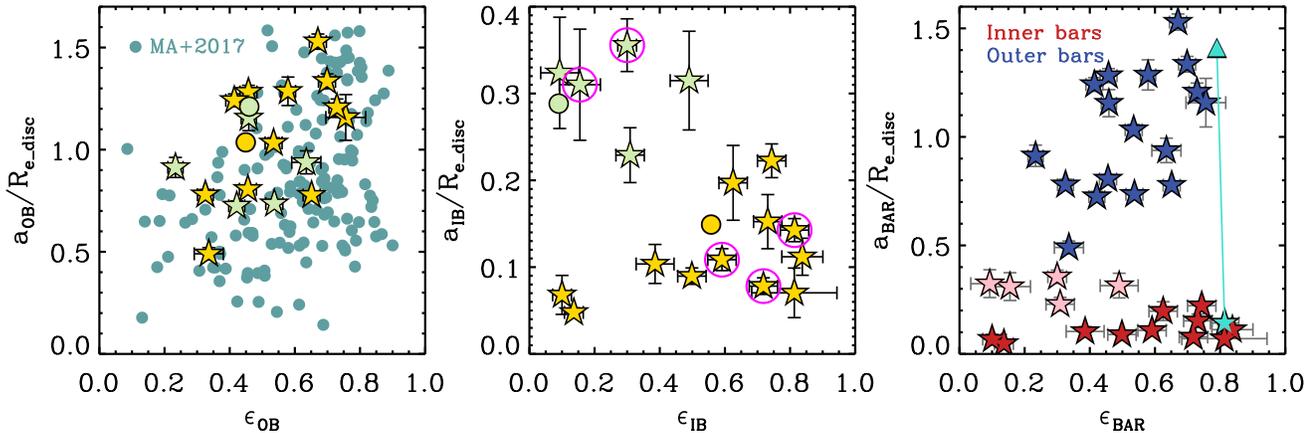}
 \caption{Deprojected bar semi-major axes (in units of the disc effective radius) versus ellipticies for the outer (left panel) 
and inner (middle panel) bars of the double-barred sample: this work (stars) and the two double-barred objects
from \citet[][circles]{MendezAbreuetal2017}. Yellow symbols represent double-barred galaxies
with $a_{\rm IB}/a_{\rm OB}<0.23$, 
whereas green symbols correspond to $a_{\rm IB}/a_{\rm OB}>0.23$.
The same properties for both inner and outer bars are shown together
in the right panel, where the long and short inner bars are coloured in pink and red, respectively, and outer bars are coloured in blue. 
Measurements of single bar semi-major axes and ellipticities from \citet{MendezAbreuetal2017} 
are included in the panel corresponding to the outer bars for the seek of comparison. While a correlation in the parameters 
for the outer bar is found, no trend is observed in the case of inner bars. The five galaxies whose inner bars have been 
spectroscopically analysed by \citet{deLorenzoCaceresetal2008,deLorenzoCaceresetal2012, deLorenzoCaceresetal2013,
deLorenzoCaceresetal2019b}
are identified with a surrounding magenta circle in the middle panel. The inner bar of one those galaxies, NGC\,5850,
is highlighted in turquoise in the right panel; it is linked with the corresponding measurement when the 
effective radius of its progenitor inner disc instead of that of the main galaxy disc is considered 
($a_{\rm IB}/R_{\rm e\_ID}$; turquoise triangle).}
\label{fig:barlengthvsell}
\end{figure*}

In Fig.\,\ref{fig:barlengthvsell} the bar length is directly compared with bar ellipticity
for the outer and inner bars of the sample. Bar lengths are provided in units of disc effective radius to prevent biases
due to different galaxy sizes. A positive correlation is found for 
outer bars 
(Spearman correlation factor $\rho=$0.49 with a significance of 0.05).
Single bars in \citet{MendezAbreuetal2017} mostly behave in an analogous way. However,
no correlation is seen for the case of inner bars, where two clouds hosting galaxies 
with different inner bar sizes appear.
This result suggests again a 
bimodal distribution of the inner bar lengths,
which lays at the basis of 
the bimodality found for the $a_{\rm IB}/a_{\rm OB}$ ratio in Fig.\,\ref{fig:lengthcomparison}
and Fig.\,\ref{fig:lengthvsht}. 

Three possibilities arise for explaining the lack of correlation within the properties of inner bars:
(a) different nature of inner and outer bars; (b) different nature of the small-scale discs
from which the inner bars are dynamically formed; and (c) different assembly stage between the two
groups of inner bars. In the following we elaborate 
each of these explanations.\\
 
(a) The correlation between bar length and ellipticity shown for outer and single bars in Fig.\,\ref{fig:barlengthvsell} 
agrees with the prediction provided by some numerical simulations that bars grow in size and strength with time
\citep[e.g.][]{SahaandMaciejewski2013,Duetal2015}. However, bar evolution is complicated as bars buckle several times
during their lives, this process affecting their overall shape and strength \citep{MartinezValpuestaandShlosman2004}.
Whatever the origin of such bar length vs. ellipticity correlation is, the fact that we do not find it for inner bars 
might be due to a totally different nature of inner bars, for which the process responsible
for the correlation in outer/single bars does not apply.
This possibility may not be fully discarded with the pieces of evidence
known so far. However, our recent observational results presented in \citet{MendezAbreuetal2019}
and \citet{deLorenzoCaceresetal2019b} do suggest that inner bars behave in a fully analogous
way than outer bars: they indeed appear to suffer one or more buckling episodes during their lives
and they are both formed dynamically from disc instabilities, the only difference being the
spatial scale of the hosting disc. A different nature of inner and outer bars does therefore seem unlikely.\\

(b) In Fig.\,\ref{fig:barlengthvsell}, 
bar lengths are normalised by the effective radius of each galaxy disc.
The size of large scale bars is known to correlate with that
of the galaxy disc as bars are the result of dynamical processes happening within the disc structure
\citep{Aguerrietal2009, Erwin2019}.
Inner bars are dynamically formed from small-scale discs \citep[hereafter called inner discs;][]{deLorenzoCaceresetal2019b}; it is thus expected
that their size correlates with that of the inner discs.
We can therefore 
suspect that the length vs. ellipticity correlation would remain for inner bars
if the effective radius of the inner disc was considered for the 
normalisation ($a_{\rm IB}/R_{\rm e\_ID}$) instead of the effective radius of the
main galaxy disc. In this case, the bimodality of the properties of inner bars would be due to
a bimodality of the properties of their progenitor inner discs. In particular, there would exist
two types of inner discs attending to how they behave with respect to their corresponding main galaxy discs.

The right panel of Fig.\,\ref{fig:barlengthvsell}
shows how the group of inner bars with $a_{\rm IB}/a_{\rm OB}>0.23$ (i.e. the longest inner bars)
lays on the length vs. ellipticity relation shaped by outer bars, whereas
the short inner bars are spread throughout the plot. 
This suggests that the inner discs hosting long inner bars are indeed related to the main galaxy disc and that is why
the correlation stands even if the inner bar size is normalised by the main disc effective radius.
This is the expected result if inner discs are formed through resonances due
to the large scale bar which, in turn, is related to the galaxy disc size.
On the contrary, the
inner discs hosting short inner bars (which lay out of the length vs. ellipticity relation)
must have a different nature since their sizes are not related with the size of the main disc.

Since the study of the faint inner discs is beyond the scope of this paper
and they have not been included in the photometric analysis, this possibility cannot be fully confirmed or
discarded for all our sample galaxies. However, a similar 2D photometric analysis with \gasp\ including
the inner discs
is presented in \citet{deLorenzoCaceresetal2019b}. For the one galaxy in common with that work, NGC\,5850,
we have overplotted the $a_{\rm IB}/R_{\rm e\_ID}$ measurement in the right panel of Fig.\,\ref{fig:barlengthvsell}.
It shows a good agreement with the bar length vs. ellipticity correlation of outer bars. 
Although this test is not conclusive for the whole sample of short inner bars, 
it provides promising support for this hypothethical scenario that can be summarised as follows: 
the bar length vs. ellipticity correlation stands for all inner and outer bars
when their size is normalised to the size of their progenitor, i.e., inner discs and main galaxy discs, respectively. 
The existence of two different kinds of inner discs explains the bimodality observed for inner bars when
the usual normalisation by the size of the main galaxy discs is applied to them.\\

(c) The third and last explanation is that inner bars are not settled structures.
If the length vs. ellipticity correlation were actually due to bars growing in length and strength
with time, it would happen only in the case of fully assembled systems which are now subject of a stable evolution.
During the bar formation process, those correlations are not expected.
Inner bars have even been found to form and dissolve one or two times before they finally
settle down \citep[e.g.][]{Wozniak2015}. 
All observational studies performed so far on double-barred galaxies agree that outer
bars form prior to inner bars (\citealt{deLorenzoCaceresetal2012, deLorenzoCaceresetal2013, deLorenzoCaceresetal2019b}),
as confirmed by their stellar populations and star formation
histories. 
Within this scenario, large-scale bars that have
developed an inner bar should be already assembled systems growing accordingly in size and strength, while
inner bars could still be immersed in  the process of settling up.
The main caveat for this explanation is that the already-mentioned observational studies
find that inner bars, although younger than outer bars, are old systems with luminosity-weighted ages
around 6\,Gyr. However, and as specifically pointed out by \citet{Wozniak2007}, the time since the dynamical
formation of the bars does not necessarily corresponds to the mean age of the bar stellar populations.
The most precise constraint for the assembly epoch of double-barred galaxies is presented in 
\citet{deLorenzoCaceresetal2019b}, where we measure that the two inner bars under study
were dynamically formed $>$4.5\,Gyr and $>$6\,Gyr ago.

The only way of reconciling this scenario with the studies about stellar populations
of double-barred galaxies is assuming that we are witnessing inner bars at different evolutionary stages.
The bimodality in the physical length of inner bars may be suggestive of
an assembled/non-assembled behaviour. 
Such possibility was already introduced by \citet{Wozniak2015}, who finds there is a first generation of
transient inner bars with small bar length ratios. These are subsequently dissolved and reformed with 
a larger ratio.
The fact that the long inner bars
lay at the bottom end of the relation for the outer bars (see right panel
of Fig.\,\ref{fig:barlengthvsell}) would be in agreement with the fact that those inner bars
correspond to already settled structures, while the remaining galaxies
would still be immersed in the process of inner bar formation.
Within this scenario, inner bars at the formation process would be shorter
than settled inner bars, in accordance with the statement
that bars grow in length with time, and inner bars would be long-lived structures as 
predicted by the numerical simulations and similarly to the case of large-scale bars.

In an attempt of exploring this last hypothesis, in Fig.\,\ref{fig:barlengthvsell}
we identify the 5 out of the 6 galaxies 
from the samples analysed in the set of papers by 
\citet{deLorenzoCaceresetal2008, deLorenzoCaceresetal2012, deLorenzoCaceresetal2013,deLorenzoCaceresetal2019b} 
and included in this analysis. In particular, the inner bar in NGC\,5850
belongs to the short-length group. This structure was formed $>$4.5\,Gyr ago as demonstrated
in \citet{deLorenzoCaceresetal2019b}. The other two galaxies
hosting short inner bars are NGC\,2859 and NGC\,4725, while NGC\,357 and NGC\,3941
have long inner bars. A stellar population analysis for all these galaxies show inner bars are 
shaped by mainly old stellar populations (mean luminosity-weighted age $\sim$6\,Gyr).
Within our proposed scenario, NGC\,5850 would host a non-assembled inner bar.
We therefore conclude  that either 
the assembly stage of the bars is not driving the observed bimodality, or the dynamical
settling of inner bars is a long process that might take several Gyr to happen.

Finally, we seek for any difference among the off-plane shape of the long and short inner bars:
three out of the seven galaxies shown in Fig.\,\ref{fig:comparisonbar3D} belong to the 
group of long inner bars with $a_{\rm IB}/a_{\rm OB}>0.23$.
No particular properties of the vertical extension of these galaxies with respect
to the remaining ones is apparent in this analysis: the clear bimodality found for the in-plane
length of inner bars does not have a counterpart in its off-plane shape.

\subsection{Will all barred galaxies be double barred at some stage?}\label{sub:discussionubiquity}

\begin{figure}
 \vspace{2pt}
 \includegraphics[angle=90., width=0.45\textwidth]{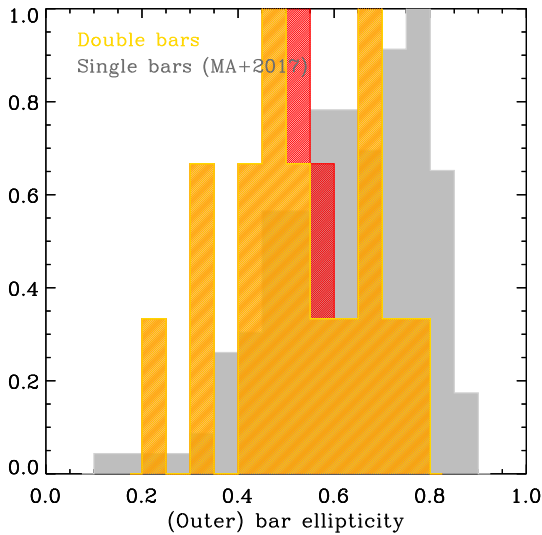}
 \caption{Normalised distribution of large-scale bar ellipticities: outer bars within the 
double-barred galaxies probed 
in this work are plotted in yellow, while the CALIFA single bars analysed in \citet{MendezAbreuetal2017}
are shown in grey. The red segments show how the distribution studied in this paper would be modified by 
including the two double-barred galaxies in \citet{MendezAbreuetal2017}.}
\label{fig:barssingledouble_ellipticity}
\end{figure}

Similarly to the debated question of whether all bars have the capability of forming
disc-like bulges through secular evolution (discussed in Paper I), it is sensible to ask whether all single-barred galaxies will 
develop an inner bar at some evolutionary stage of their lives. This hypothesis
backs on the two possible formation scenarios presented for inner bars.
\citet{FriedliandMartinet93, Helleretal2001, SahaandMaciejewski2013},
among others, obtain that outer bars are formed first and the material flown through them
is responsible for creating the inner bar structures thanks to the dynamics
of the outer bar. \citet{DebattistaandShen2007} and \citet{Duetal2015}, on the other hand, 
find that inner bars are formed dynamically from cold inner discs. Although this scenario does
not invoke the presence of gas, the most likely possibility is that the inner discs are formed in 
a star-forming process out of gas that has inflown along the outer bar. 
We remark again that our stellar population analyses of double-barred galaxies presented in
\citet{deLorenzoCaceresetal2012,deLorenzoCaceresetal2019b}
support the dynamical origin of inner bars.
But regardless of which scenario dominates the inner bar formation, 
the premise is that, as long as gas is present, inner bars could be formed, 
either directly or through the gas-rich formation of an inner disc.

Assuming once again that large-scale bars grow with time
since their assembly (see previous discussion about
Fig.\,\ref{fig:barlengthvsell}), our aim here is to check if bars hosting inner bars are more evolved
than pure single bars.
If (outer/single) bars are long-lived, 
the observables
indicating the time since the assembly of the bar, i.e. length and ellipticity
(as proxy for strength), should acquire larger values in the case of 
outer bars (that have developed an inner bar inside) than in the case of single bars
(that have not formed inner bars yet).

Figure \ref{fig:barssingledouble} compares the sizes of truly single-barred galaxies
from \citet{MendezAbreuetal2017} with the outer bars
of our double-barred sample. Outer bars are not systematically longer than single bars.
We complete this comparison with a similar plot for the ellipticities of large-scale
bars in single- and double-barred galaxies, shown in Fig.\,\ref{fig:barssingledouble_ellipticity}.
Notwithstanding the fact that both distributions are different as proven with a Kolmogorov-Smirnov test
(with a probability of being similar distributions of 0.05), 
the derived trend is the opposite than expected, with truly single bars
being stronger (i.e. with higher ellipticity values) than outer bars.

Although our analysis does not support the ubiquity of inner bars, this possibility
cannot be ruled out yet as there are several caveats hampering these results:
first, the presence of gas is required for either forming the inner bar or, most likely, for forming
the inner disc which will develop the inner bar. The presence of gas is not considered in this analysis.
Second, other galaxy properties 
(probed for example with the Hubble galaxy type) besides the dynamical age of
the outer bar may be influencing the moment at which the inner bar is formed. 
And third, bars may actually not grow in length and strength in a stable way with time, as discussed
in previous Sect.\,\ref{sub:discussionsettle}.
Our sample of 17 galaxies is not large enough to probe more parameters and provide
robust conclusions but it is worthwile to keep this possibility in mind.


\section[]{Conclusions}\label{sec:conclusions}
Following the thorough photometric analysis of double-barred galaxies presented 
in Paper I,
we study here the intrinsic photometric properties of inner and outer bars.
They are furthermore compared with widely used results from analyses of integrated
images and discussed within the context of the formation and evolution of these complex galaxies.

The main observational pieces of evidence observed here are:
\begin{itemize}

\item The 17 inner bars analysed in this work show lengths between 0.17\,kpc and 2.4\,kpc (in-plane semi-major axes). 

\item Inner bars may be longer than outer bars hosted by other double-barred systems,
as we find outer bars spanning from 1.9 \,kpc to 22.5\,kpc in length (in-plane semi-major axes).

\item A bimodal distribution of the length ratio between inner and outer bars is found, which is in turn
related to two distinct groups of bars in the inner bar length vs. ellipticity diagram.
The length ratios are of 11\% \citep[similar to the measurement by ][]{ErwinandSparke2002}
and 35\%. The origin of this behaviour is related neither to the Hubble type of the
host galaxy nor to the bulge properties. No equivalent
bimodality in the off-plane thickness is found.

\item No preferred orientation between the outer and inner bars is found,
as probed by their deprojected position angles.

\item There is no relation between the relative orientation of the two bars and 
the length ratio.

\item No particular trend is found between the strength of the inner and outer 
bars, being the inner bar weaker or stronger than the outer bar.
The bar strength has been probed using the in-plane ellipticity and off-plane thickness
as proxies.

\item Length and ellipticity are correlated for the case of outer bars, 
as expected for a fully assembled, settled up structure.
Although such correlation is not found for inner bars in general, the group
of long inner bars does lay in the short end of the relation for outer bars.

\end{itemize}

The bimodality observed for the length of the inner bars may be a consequence of
(a) a different formation or evolution path for some inner bars; (b)
a different nature of the inner discs from which the inner bars are dynamically formed; 
or (c) a different assembly stage at the moment we are witnessing these galaxies.
Previous spectroscopic studies of the stellar populations and star formation
histories of double-barred galaxies have demonstrated that inner bars
form and behave in an analogous way to large scale bars, thus suggesting
that option (a) is very unlikely. These studies have used inner bars belonging 
to the two subgoups found in this work: the long and the short inner bars. 
Both explanations (b) and (c) could lay at the basis of the bimodality, but we note
that a different assembly stage (c) would imply that the dynamical assembly 
of inner bars is a slow process that may take several Gyr.

We have also explored whether there is evidence of outer bars being
more evolved systems than single bars. If so, this would leave room for the possibility
that all bars, once they have lived enough, will develop an inner bar 
at a later stage of their lives. However, outer bars within double-barred systems
do not appear longer or stronger than purely single bars.\\

The photometric results presented here in combination with
a detailed spectroscopic study of the stellar populations and star formation histories of 
double-barred galaxies
is the most powerful strategy to assess the formation of these systems. 
Such combination of techniques has already been sucessfully used in \citet{deLorenzoCaceresetal2019b}
for a limited sample of two individuals. 
Its application to a large sample of double-barred galaxies, as that studied in this paper, 
would be necessary to constrain the nature of inner bars and their stability over time.

\begin{onecolumn}
\begin{table}
\caption{Deprojected $r'$-band physical parameters for the double-barred galaxies.}
\begin{center}
\begin{tabular}{l|c|cccc|ccccc}
\hline
Galaxy     &  R$_{\rm e\_disc}$ (kpc) & $a_{\rm IB}$ (kpc) & $a_{\rm e\_IB}$ (kpc) & $\epsilon_{\rm IB}$ & PA$_{\rm IB}$ ($^{\circ}$) &  $a_{\rm OB}$ (kpc) & $a_{\rm e\_OB}$ (kpc) & $\epsilon_{\rm OB}$ & PA$_{\rm OB}$ ($^{\circ}$) \\
(1) & (2) & (3) & (4) & (5) & (6) & (7) & (8) & (9) & (10) \\
\hline
\hline
NGC357   &  7.64  $\pm$  0.36  &  2.37  &  0.90  &  0.15  &  -87.31  &  7.17   &   2.71  &   0.64  &    -62.84 \\
NGC718   &  4.18  $\pm$  0.16  &  0.43  &  0.18  &  0.39  &  16.66   &  5.37   &   1.85  &   0.58  &    -31.12 \\
NGC2642  &  12.78 $\pm$  0.87  &  0.90  &  0.38  &  0.81  &  -19.66  &  14.80  &   5.59  &   0.76  &    116.14 \\
NGC2681  &  2.50  $\pm$  0.06  &  0.17  &  0.06  &  0.10  &  -148.95 &  1.95   &   0.74  &   0.33  &    -98.43 \\
NGC2859  &  11.54 $\pm$  0.19  &  1.25  &  0.47  &  0.59  &  57.25   &  9.31   &   3.52  &   0.46  &    -16.22 \\
NGC2950  &  3.76  $\pm$  0.06  &  0.74  &  0.25  &  0.63  &  66.47   &  4.82   &   1.82  &   0.46  &    -0.29  \\
NGC2962  &  7.14  $\pm$  0.27  &  2.25  &  0.85  &  0.49  &  -87.72  &  6.52   &   2.76  &   0.23  &    -26.31 \\
NGC3368  &  7.79  $\pm$  0.13  &  0.70  &  0.26  &  0.50  &  115.76  &  9.67   &   3.65  &   0.41  &    102.80 \\
NGC3941  &  3.13  $\pm$  0.08  &  1.11  &  0.42  &  0.30  &  -105.66 &  2.27   &   0.96  &   0.42  &    -25.36 \\
NGC3945  &  6.44  $\pm$  0.11  &  0.98  &  0.37  &  0.73  &  75.08   &  6.65   &   2.82  &   0.53  &    70.55  \\
NGC4314  &  6.50  $\pm$  0.16  &  0.31  &  0.11  &  0.14  &  -9.37   &  7.85   &   2.97  &   0.73  &    -32.12 \\
NGC4340  &  3.44  $\pm$  0.06  &  0.77  &  0.26  &  0.74  &  13.38   &  5.27   &   2.59  &   0.67  &    20.05  \\
NGC4503  &  3.80  $\pm$  0.06  &  0.87  &  0.39  &  0.31  &  -92.81  &  2.80   &   1.19  &   0.54  &    -44.63 \\
NGC4725  &  11.12 $\pm$  0.18  &  0.87  &  0.33  &  0.72  &  -45.03  &  8.65   &   3.27  &   0.65  &    -132.63\\
NGC5850  &  16.87 $\pm$  0.28  &  2.41  &  1.02  &  0.81  &  46.95   &  22.52  &   7.75  &   0.70  &    108.67 \\
NGC7280  &  5.31  $\pm$  0.25  &  0.59  &  0.25  &  0.84  &  -52.75  &  2.60   &   1.10  &   0.34  &    36.72  \\
NGC7716  &  4.53  $\pm$  0.17  &  1.47  &  0.55  &  0.09  &  -80.95  &  5.23   &   2.22  &   0.46  &    9.88   \\
\hline
\end{tabular}
\end{center}
\label{tab:deprovalues}
Notes. (1) Galaxy name; (2) physical disc effective radius in kpc; (3) deprojected physical inner bar length (semi-major axis) in kpc;
(4) deprojected inner bar effective radius in kpc; (5) deprojected inner bar ellipticity; (6) deprojected inner bar position angle in degrees;
(7) deprojected physical outer bar length (semi-major axis) in kpc; (8) deprojected outer bar effective radius in kpc; 
(9) deprojected outer bar ellipticity; and (10) deprojected outer bar position angle in degrees.
Deprojection adds its own uncertainties; they depend on the galaxy inclination but are generally within 10\% of the values. 
We refer the reader to
\citet{Zouetal2014} for an analysis on the uncertainties caused by deprojection over bar parameters. \end{table}
\end{onecolumn}
\twocolumn

\section*{Acknowledgments}
The authors are very grateful to the anonymous referee of this paper for his/her
careful revision and interesting suggestions.
AdLC acknowledges support from grants ST/J001651/1 (UK Science and Technology
Facilities Council - STFC) and AYA2016-77237-C3-1-P (Spanish Ministry of Economy and 
Competitiveness - MINECO).
JMA acknowledges support from the Spanish Ministry of Economy and Competitiveness (MINECO) 
by grant AYA2017-83204-P.
LC acknowledges financial support from Comunidad de Madrid under Atracci\'on de Talento 
grant 2018-T2/TIC-11612 and the Spanish Ministerio de Ciencia, Innovaci\'on y Universidades 
through grant PGC2018-093499-B-I00.

\bibliographystyle{mn2e}

\label{lastpage}

\end{document}